\documentstyle[psfig,twoside ,fleqn,espcrc2]{article}

\newcommand{\AmS}{{\protect\the\textfont2
  A\kern-.1667em\lower.5ex\hbox{M}\kern-.125emS}}

\title{Clear Evidence of a Continuum Theory of 4D Euclidean Simplicial Quantum Gravity 
\thanks{presented by S.Horata}}

\author{H.S.Egawa
        \address{Department of Physics, Tokai University, 
        Hiratsuka, Kanagawa 259-1292, Japan}
        $^{,}$
        \address{Theory Division, Institute of Particle 
        and Nuclear Studies, KEK, High Energy Accelerator 
        Research Organization, Tsukuba, Ibaraki 305-0801, Japan}
		,
        S.Horata$^{\,\, {\rm b}}$
        and 
        T.Yukawa 
        \address{Coordination Center for Research and Education, 
        The Graduate University for Advanced Studies, 
        Hayama, Miura, Kanagawa 240-0193, Japan}
        $^{\! , \,\,{\rm b}}$}
\begin{document}

\begin{abstract}
Four-dimensional (4D) simplicial quantum gravity coupled to both
scalar fields ($N_X$) and gauge fields ($N_A$) has been studied using
Monte-Carlo simulations. 
The matter dependence of the string susceptibility exponent
$\gamma^{(4)}$ is estimated.
Furthermore, we compare our numerical results with
Background-Metric-Independent (BMI) formulation conjectured to describe the
quantum field theory of gravity in 4D.
The numerical results suggest that the 4D simplicial quantum gravity is 
related to the conformal gravity in 4D.
Therefore, we propose a phase structure in detail with adding both scalar
and gauge fields and discuss the possibility and the property of a
continuum theory of 4D Euclidean simplicial quantum gravity.
\end{abstract}
\maketitle 
\section{Introduction} 
\vspace{0.3cm}
4D Euclidean Simplicial Quantum Gravity (4D Eucl. SQG) has been
investigated from several points of view.
As a first step, the phase diagram in case of pure gravity has been
studied numerically.
Especially, it has been shown that the phase structure in case of
pure gravity has two distinct phases, the crumpled phase and the
elongated phase, and the phase transition between the two phases is
shown to 1st order.
Moreover, the string susceptibility exponent ($\gamma^{(4)}$) takes
positive value at the critical point.
As a result, it is difficult to construct to a continuum theory.

For the next stage, the case of adding gauge matter fields has been
intensively investigated\cite{BBKPTT,EFHTY}.
Then, the phase structure changes drastically with adding gauge
matter fields.
the smooth (crinkled) phase with negative $\gamma^{(4)}$ was
found even for $N_{A} \ge 1$\cite{EFHTY}.
And the phase transition between the crumpled phase and the smooth phase
has been shown to be continuous\cite{EHTY}.
We consider that the dynamical triangulated manifold is stable with adding
some matter fields.
It is expected that the 4D Eucl. SQG coupled to matter fields reaches to
the continuum theory.
%
%

Numerical results indicate that the 4D Space-time becomes stable with
adding the matter fields.
It is supported by the analytical calculations of 4D conformal
gravity\cite{AMM,HS,HAMA}.
We emphasize that the analytical calculations from 4D conformal filed
theory are similar to numerical ones in case of adding matter fields.
In this article, we would like to clarify the relation between 4D
conformal field theory and 4D Eucl. SQG, quantitatively.

We consider that $\gamma^{(4)}$, as well as $\gamma^{(2)}$ in 2D,
is given by a function of the number of matter fields,
\begin{equation}
 \gamma^{(4)} = \gamma^{(4)}(N_{X} + 62 N_{A}).
\end{equation}
Coefficients for the number of fields relates to the central charge of
the model and can be calculated exactly from 4D conformal field theory.
In order to investigate the relation between 4D Eucl. SQG and
the quantum field theory of gravity, we estimate the matter dependence for
$\gamma^{(4)}$ by numerical simulation with adding scalar fields ($N_X$)
and gauge fields ($N_A$).
%
%
\section{Model}
%
The action $S$ is given on the 4D simplicial manifold as,
\begin{equation}
 S = S_{G} + S_{A} + S_{X},
\end{equation}
where $S_{G}$, $S_{A}$ and $S_{X}$ denote the action of the gravity
part, the U(1) gauge fields ($A$) and scalar fields ($X$), respectively.
For the gravity part, we use the discritized Einstein-Hilbert action in 4D,
\begin{equation}
 S_{EH}[\kappa_{2},\kappa_{4}]=
\kappa_{4}N_{4}-\kappa_{2}N_{2}, 
\end{equation}
where $N_{i}$ denotes the number of $i$-simplex.
The two parameters, $\kappa_2$ and $\kappa_4$, correspond to the inverse
of the gravitational constant and the cosmological constant,
respectively.
The action,
\begin{equation}
 S_A=\sum_{t_{ijk}} o(t_{ijk}) (A_{ij} + A_{jk} + A_{ki})^2,
\end{equation}
corresponds to that of gauge matter fields, where $A_{ij}$ is the
non-compact gauge field defined on the link $l$ between vertices $i$
and $j$ with the sign convention, $A_{ij}=-A_{ji}$.
The weight factor $o(t_{ijk})$ is given by the number of 4-simplices
sharing the triangle $t_{ijk}$.
We also add massless scalar fields,
\begin{equation}
 S_{X}=\sum_{ij}(X_i - X_j)^{2},
\end{equation}
where $X_i$ is the scalar field defined on the vertex $i$.
Then the partition function is given as,
\begin{eqnarray}
 Z(\kappa_{2},\kappa_{4},N_A,N_X) = \sum_{T}
  e^{-S_{G}(\kappa_{2},\kappa_{4})} \qquad\qquad\quad& & \nonumber \\
\prod_{N_A,N_X}
  \left(
   \int \prod_{l \in T} dA_l e^{-S_{A}(A_l)} 
   \int \prod_{i \in T} dX_i e^{-S_{X}(X_i)} 
 \right).
\end{eqnarray}
We sum over all 4D simplicial triangulation ($T$) in order to carry out 
a path integral about the metric. 
Here, we fix the topology with $S^{4}$.
Measurements are made at every 100 sweeps.

\section{Matter Dependence of 4D Simplicial Quantum Gravity}
\begin{figure}
\centerline{\psfig{file=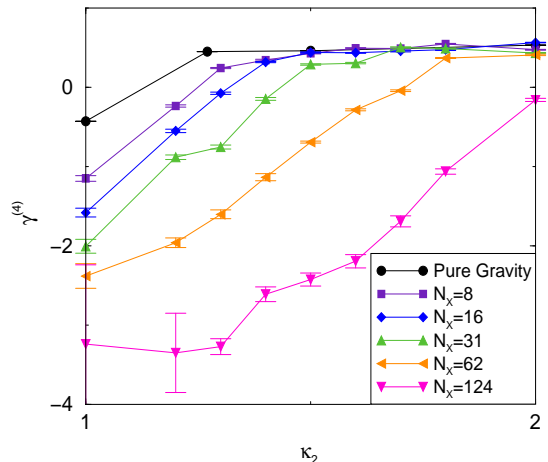,height=7cm,width=7cm}} 
\vspace{-12mm}
\caption
{
 The string susceptibility exponent $\gamma^{(4)}$ plotted for various
 numbers of scalar fields ($N_X$) versus the coupling constant ($\kappa_2$).
}
\vspace{-0.7cm}
\label{GAMMA}
\end{figure}

In this section, we report on the matter dependence of $\gamma^{(4)}$.
We measure $\gamma^{(4)}$ by using the MINBU method.
In Fig.\ref{GAMMA}, we plot $\gamma^{(4)}$ for various numbers of scalar
fields versus $\kappa_2$.
%
%
It is clear that $\gamma^{(4)}$ becomes negative with adding scalar
fields, which is similar to the case of adding gauge fields.

\begin{figure}
\centerline{\psfig{file=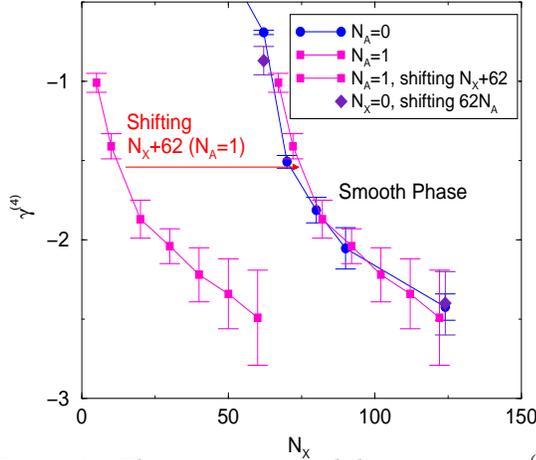,height=7cm,width=7cm}} 
\vspace{-13mm}
\caption
{
 The string susceptibility exponent $\gamma^{(4)}$ plotted versus
 numbers of scalar fields ($N_X$).
}
\vspace{-7mm}
\label{MATTERDEP}
\end{figure}

Then, the matter dependence of $\gamma^{(4)}$ is shown.
In Fig.\ref{MATTERDEP}, we plot $\gamma^{(4)}$ versus the numbers of
scalar fields ($N_X$).
%
%
In order to see the matter dependence of $\gamma^{(4)}$, we compare two
different cases which are the case of no scalar fields ($N_{A} = 0$) and
the case of adding one gauge field ($N_{A} = 1$).
For both of these cases, we measure $\gamma^{(4)}$ for various numbers
of scalar fields and plot both of the results for $N_{A} = 0$ and $N_{A}
= 1$, respectively in Fig.\ref{MATTERDEP}.
As compared both of the results, the case of $N_{A} = 1$ can be
corresponded to that of  $N_{A} = 0$ with shifting $N_{X} \rightarrow
N_{X} + 62 \times 1$.
Furthermore, we also plot the results of $N_{A} = 1$ and $N_{A} = 2$
with no scalar field $N_{X} = 0$.
Both of these results are placed on the results of $N_A = 0$ with
varying numbers of scalar fields ($N_X$).

Our numerical results suggest that $\gamma^{(4)}$ is the function of the
number of scalar fields ($N_X$) and gauge fields ($N_A$),
\begin{equation}
 \gamma^{(4)} = \gamma^{(4)}(N_X + 62 N_A).
\end{equation}
Moreover, the coefficient 62 corresponds to the result of 4D conformal field
theory.
%

\section{Summary and Discussions}
Let us summarize and discuss the main points made in the previous section.
We measure $\gamma^{4}$ for various numbers of scalar fields ($N_X$).
In Fig.\ref{MATTERDEP}, we show that the matter dependence of
$\gamma^{(4)}$ is given as,
$
 \gamma^{(4)} = \gamma^{(4)}(N_X + 62 N_A).
$
The coefficients $62$ is equivalent to the result of 4D conformal field
theory.
Numerical results suggest that 4D Eucl. SQG is related to the conformal
gravity in 4D.

Furthermore, we compare our numerical results with
BMI formulation to describe the quantum field theory on a non-dynamical
background metric theory of gravity with the dynamics of the
traceless mode as well as the conformal mode in 4D\cite{HS,HAMA}.

\begin{figure}
\centerline{\psfig{file=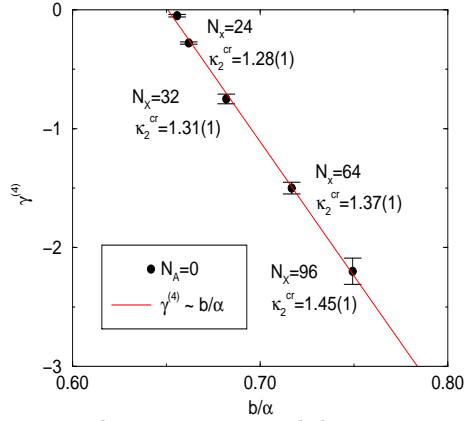,height=6cm,width=6cm}}
\vspace{-13mm} 
\caption
{
 The string susceptibility exponent $\gamma^{(4)}$ plotted versus
 $b/\alpha$. 
}
\vspace{-7mm}
\label{CRITICALGAMMA}
\end{figure}

In order to compare the results of BMI formulation with 4D Eucl. SQG,
we focus the matter dependence of $\gamma^{(4)}$.
From BMI formulation, $\gamma^{(4)}_{BMI}$ can be estimated as,
\begin{equation}
 \gamma^{(4)}_{BMI} \propto \frac{b}{\alpha},
\end{equation}
with the ambiguity of the volume definition.
The quantity $b$ plays the role of the central charge in BMI
formulation and $\alpha$ is the dimension of the cosmological constant.
We show the numerical result at the near critical point in
Fig.\ref{CRITICALGAMMA}

%
Numerical results show the fact that $\gamma^{(4)}$ is proportional to
$b/\alpha$.
It corresponds to the estimation from BMI formulation.
Then, $\gamma^{(4)}$ is given as the function of the quantity $b$
that plays the role of the central charge in 4D.
It is the same situation as $\gamma^{(2)}$ is given as the function of
the central charge $c$ in 2D.

Numerical results suggest that 
4D Eucl. SQG coupled matter fields relate to the conformal invariance
in 4D and BMI formulation.
It should be clear that a continuum 
theory of 4D Eucl. SQG depends on dynamics of
the traceless and conformal mode.
Moreover, we expect that the properties of 4D quantum gravity are
similar to those of 2D quantum gravity.


\end{document}